\documentclass[12pt,thmsa]{article}
\usepackage{sw20aip}

\input{tcilatex}
\begin{document}

\title{THE DECREASE IN PHYSICS ENROLLMENT }
\author{Samina S. MASOOD \\
Department of Mathematics, School of Arts and Science, \\
SUNY Institute of Technology, Utica, NY 13504\\
Email: masoods@sunyit.edu}
\maketitle

\begin{abstract}
We discuss and analyze the fact that physics is generally losing its ability
to captivate students who may possess the potential to enhance the quality
of our future in this age of technology. We have tried to investigate the
reasons behind this low enrollment in the light of the results of a few
surveys with the undergraduate students in different physics courses and in
current relevant college programs. It is not an exclusively descriptive
issue, so our analysis is a way to delineate the details of the matter
leading to the suggestions for future improvements.
\end{abstract}

\section{INTRODUCTION}

It is now well known that physics is losing its hook on the popular choices
of elective undergraduate courses. The gradual decrease in physics
enrollment is a thought-provoking issue for us all. Physicists are all
concerned about the future of physics in this age of technology. But this
problem is even more serious in the US where the great development in
technology has been achieved through the continuous efforts of a few
decades. Now, if this trend of low enrollment persists, most of the relevant
physics programs may fade out and as a consequence, we may have to
compromise standards, probably not the best thing to do. As a reality,
students do not prefer to enroll in physics in the presence of other
options. With increasing technological demands, though, it should be the
other way around. So, we should accept the responsibility of failing to
provide compelling reason to learn physics.

Nowadays, students will register in physics courses only if they have to
take it to fulfill the mandatory needs of their major programs. In this
situation, the physics undergraduate major and even independent physics
departments are already disappearing from the community colleges. We would
like to discuss here the root cause of this issue and point out those
factors that lead to this situation. Avoiding these causes one by one should
be considered as a part of the solution to the problem. We look at this
issue based on our discussions with students at different levels from more
than fifteen years of teaching and our personal experience of working with
the physics community in different continents at different levels for more
than two decades. We understand that we have to think about this issue very
carefully without ignoring minor details. Some research is being done on a
few related topics but, according to our information, this topic has not yet
been discussed thoroughly elsewhere. In the reference section we have
mentioned a few references [1-5] for the interested readers for further
study of some of the individual topics. In the next section we give an
overview of this problem in detail and point out the basic reasons for the
decline in enrollment. In section III we discuss a few of the important
factors in more detail. In sections IV, V and VI more individual issues are
discussed. The last section VII emphasizes some of our immediate
suggestions. This discussion will include the details of the underlying
problems and procedures to get rid of these causes as a solution. Especially
sensitive topics are briefly mentioned for completeness. The authors request
readers not to take it personally.

\section{OVERVIEW\ OF\ THE\ PROBLEM}

When we start thinking about the reasons for the low enrollment, it appears
to be a very tricky question. We cannot simply describe the reasons behind
it. We need to analyze it in the complete scenario of educational system,
employment opportunities, teaching and research needs and future
perspectives. But we have to pause for a moment and ask ourselves a question
before this. Do we all really care so much about it? Usually, we don't.
Probably we are convinced that we just need trained people in the field of
technology and not basic sciences. It is probably because we just need very
few experts in basic sciences and a large number of engineers and
technicians to work for the development of technology. Several technologists
can work with scientists even without any specialized high level science
background. This is true. Data analysis and other computational help are
examples. They could be handled without very strong science background to
start with, as long as you are working in an active research atmosphere and
have expert help available. However, we still need physicists to propose the
basic design and purpose of the experiment. They can check the feasibility
of certain features of the designs of new projects and contribute to the
interpretation of the experimental results. Therefore, it is necessary to
keep on producing quality physicists for our future needs. There should be
much larger number of physics majors so that adequate physics training could
be provided to enough future technologists. And we could even produce a
sufficient number of competent physicists to fulfill future needs. Obviously
we do not want to run the higher level physics education programs mostly for
the foreign students. Available data [6] indicates the growing enrollment of
foreign students at the undergraduate level. However, the ratio of foreigner
graduate students in physics is increasing even more. To avoid this, we need
to have physics majors in most of the colleges, in addition to the research
universities. Otherwise, we might face more serious consequences in the
future and almost limit the physics programs to the research universities
and the institutes of technology.

We should start this analysis by convincing ourselves that physics is a very
interesting subject. Physics is one of the very important undergraduate
majors for most of the specialized engineering and technology programs. Let
us look at the reasons why physics should be considered a good undergraduate
major in support of most of the technical programs.

\begin{enumerate}
\item  Physics is a very useful basic science subject and has to be taken as
a basic requirement for several undergraduate/graduate degree programs in
Science and Engineering.

\item  Some very important issues in Imaging Science, Astronomy, Cosmology
and Space Sciences can only be understood in the light of physics.

\item  Physics knowledge is sometimes crucial in designing the projects in
environmental science, medical diagnostics, defense, and many other areas
related to the improvement of human security and livelihood. Especially, the
working of all type of equipments can better be understood with physics
background.

\item  Development in modern technology needs multi-disciplinary
contributions and physics is indispensable in this regard.

\item  Fundamental research in physics, undoubtedly, has its own importance
and applications. Findings of basic research help in determining new
directions of development in technology. There is still so much to be done
in the field of physics as in any other field of science.

\item  Good physics teaching is essential to produce good physics teachers,
simply to teach undergraduate courses in the future.
\end{enumerate}

Moreover, I personally believe that physics plays a constructive role in the
development of personalities and societies. Especially, in the current
period of scientific and technological development, physics knowledge is
very helpful in improving daily life. It even helps in self-reliance, human
safety and money savings. Some of the common issues are safety tips in
driving, energy and fuel conservation, the use and maintenance of appliances
and several other related things. Physics labs give some hands-on experience
in equipment handling and attract students to physics. Dr. Richard Hake has
discussed the effectiveness of interactive engagements based on an extensive
survey [7]. Physics courses could be useful for minor repairs at home even
without any specific training. Moreover, the effect of physics on
personality and the approach towards life is not ignorable. Science courses
help to develop the scientific approach in our daily life and make us more
reasonable and practical persons. So in general, physics helps to improve
the quality of daily life. These are known facts but we have to communicate
it to the students in an appropriate way so that they can realize the
general importance of this subject. One of the options is to introduce
interesting inter-disciplinary courses to increase the popularity of
science, especially physics and its applications to daily life. Some of the
already existing similar courses could be introduced in other colleges also.

It is also worth mentioning at this stage that our topic is physics in
general, which includes both fundamental physics and applied physics.
However, applied physics can only develop in the presence of good knowledge
of fundamental physics, either theoretical or experimental. In applied
physics, the basic concepts of physics are used to develop application of
physics in different directions. Applied physics directly helps in the
development of engineering and technology. This division includes all
branches of physics, depending on the demands and requirements. All the
branches of physics have a lot of overlap among themselves. This discussion
is not in the scope of our topic at the moment.

In fact, almost every good physics department has to have a couple of very
good groups in theoretical physics to work on the basic physics issues and
also to help the big experimental groups whenever they are needed. All the
big experimental labs also have a very strong theory division/department.
Theorists do independent theoretical or mathematical research work and also
can suggest new experiments based on their research. They are not confined
to the analysis of the final results or helping experimentalists in
designing the experiment or interpreting the results. They suggest and help
in designing new experiments also. In addition to that, they can teach a
great variety of the basic physics courses with their broad based
theoretical background. We are personally convinced that the theorist should
teach the experimentalists so that they can have some exposure to the
theoretical / mathematical approach also. Similarly, for the theorists it is
always better to have some exposure to the experimental approach, though
they get it through teaching labs to some extent. Basic physics knowledge
with a mathematical approach could be very helpful for engineers and
technologists. In this situation, physics should obviously be considered a
great second undergraduate major for future engineers. This even helps to
develop interdisciplinary courses very well. With all this relevance with
modern educational programs, the lack of interest in the subject is strange
and unnatural. We now have to find out the reasons why students are losing
their interest in this subject, leading to low enrollment in physics courses.

We will also like to mention that our analysis is mainly based on the
community colleges, state colleges and small institutes that only offer a
limited number of undergrad or graduate programs. It also includes surveys
from an institute of technology. However, this survey and analysis may not
be so true for the research universities or institutes with advanced physics
programs. They have a much better funding situation and much better
equipment. In these institutes, active researchers are teaching elementary
level physics courses with low teaching load and less job dependence on
instructor evaluation. They have help in grading and have graduate students
for the individual support of the undergraduate students in elementary
physics courses. In this situation, they can afford to have large lecture
sections. Teachers do not have an excessive load of individual guidance
during their office hours. They have broad-based knowledge of physics and
can be impressive and effective teachers and can become role models for
students because of their academic standing. They have the liberty to try
new effective teaching tools [8] and do not have to compromise standards for
job security. Their job security depends primarily on their participation in
research activities and research funding. So, they have a much better
situation and are still able to manage research programs. This difference in
situation supports our analysis below. But these places are usually
expensive places and most students cannot afford to go to these institutes,
especially during the first couple of years of college. This is the very
time when most decide their goals and future profession. Therefore, in the
future physics majors will be coming mostly from better financial
backgrounds and from expensive colleges only.

However, it is also noticed that in the community colleges or even sometimes
in the state colleges, less qualified teachers are preferred to more
qualified ones and these colleges tend to discourage research also.
Moreover, engineers at some places are teaching the physics. We personally
have seen cases where engineers are preferred over the physics doctorate
degree holders for physics teaching because the recruiters were not physics
degree holders themselves. Here standards are obviously difficult to
maintain. A similar kind of effect on the quality of teaching could be
observed if a theoretical physicist with a limited knowledge of engineering
applications is teaching engineering courses. So it is just a question of
having more qualification and broader background in the course to be taught.

In the next section we will mention the reasons for low enrollment in a
little more detail. We have also incorporated accumulated data based on
several surveys from different colleges in New York State that were
performed during the past several years. Some of this analysis is based on
direct experience of teaching at different levels starting from very
elementary level physics courses to the highly specialized Ph.D. courses.
During this time, we directly worked with students from different cultural
and ethnic background and we were exposed to the teaching systems of a few
different countries. It is also worth mentioning that we have not included a
quantitative analysis of this data in this discussion, though our
conclusions are partly drawn from this data.

\section{LOSS OF INTEREST IN PHYSICS}

The first reason is obviously the image of the subject itself. It is
undoubtedly a subject that needs adequate math background and dedication.
The math component of the subject is a big repelling factor for students.
Especially, non-science majors do not like to take physics because it needs
much more involved math at every level, compared to other courses. I have
taught astronomy courses to non-science majors and have seen this difference
a lot. Yet, astronomy is a popular course because it can be taught to
students without substantial math background. Students, even after finishing
the pre-requisite math courses, have difficulty using calculators for very
simple numerical calculations. They do not like to play with numbers at all.
At least 70\% of the non-science majors in astronomy courses do not have the
concept of units and knowledge of scientific notation. They learn to plug in
numbers in a formula but still are not able to evaluate the result. Though
first math course in college is always a pre-requisite. Use of the
calculator is a difficult task for them, though they are expected to know
this, as they are not supposed to be able to pass pre-requisite math courses
without this skill. Keeping track of all the units or other minor details of
calculations is not even mentionable. These students cannot be expected to
feel comfortable in the first algebra based course of physics. In this
situation, physics courses with minimum math component are very mathematical
courses for them even if they have at least covered the prerequisites. They
may, for example, consider conceptual physics, physical science and/or
introductory astronomy as very mathematical courses. But, I think the
minimum required math background for physics courses is what is actually
needed to improve the quality of daily life. Math itself is not a very
popular subject either, though this discussion is beyond our current topic.
Math majors do not have to take physics because physics is not usually
required for math majors. However, mathematical concepts are generally
considered to be difficult to understand for non-math majors. Understanding
the concepts of physics with the help of mathematical equations is even more
challenging. Students prefer to give it up without even trying. Apparently,
this subject looks more demanding and less interesting. Students think that
they will have to spend a lot of their study time to pass the course and
would not be able to get good grades out of it, either. They don't even find
it interesting. In this situation, it becomes a complicated problem. So we
need to make it sufficiently attractive in both ways to get better
enrollment in physics. We see different types of students in different level
physics courses. It is also true that we need dedicated students in physics
and only such students can survive in the higher-level courses. Regardless
of the very low number of physics majors, our discussion is not about the
physics majors. However, non-science majors mostly prefer other science
courses to physics. Let us first look at the reasons that why physics is
losing its popularity among the undergraduate students and cannot attract
many promising students.

After these introductory remarks, we can include a somewhat quantitative
analysis based on student surveys. We tried to collect some data from the
undergraduate students to find out that what they think about it. Here is
data collected from 300 randomly picked undergraduate students in different
science and non-science programs in different colleges in New York State
altogether. Almost half of these students were science majors. It includes
community colleges, state colleges and a private institute of technology. We
are not going to analyze this data individually at the moment. Also note
that the following figures refer to students who had not yet taken physics.
When some of these students did study physics their opinion was
significantly changed. We are still in the process of data collection and
would like to report the detailed quantitative analysis of this study later.
The overall results from the students who have never studied physics courses
in a college are as follows:

\begin{enumerate}
\item  Around 80\% of the students think that physics is a very difficult
mathematical subject. And most of these students had inadequate math
background.

\item  40\% have no interest in physics and consider it a useless and boring
subject.

\item  50\% of the registered students want to pass physics courses to
fulfill the requirement of the degree but do not like physics.

\item  70\% students of the undergrad algebra based physics class (including
around 20\% non-science majors) think that this subject will consume most of
their time outside a classroom and still not yield good grades. It could
even have a bad effect on their other grades. Some of them even liked it as
a subject or liked the teacher.
\end{enumerate}

Some randomly picked students were asked a common question during
discussions on what is the use of physics. For them, it is a pretty abstract
subject that teaches the applications of mathematical equations. They cannot
find any reason for studying physics and its connections to daily life. Most
of these students never even try to take physics so this image may be from
their exposure to physics from high schools and from their friends who took
physics at some level. These results are not surprising for physics
teachers, because we usually come across similar remarks. Almost every
school has a more-or-less similar type of situation, though the colleges
with two-year physics programs and less qualified teachers have even higher
numbers of the above comments. This leads to further decrease in the physics
enrollment as well as a job cut in physics. As a natural consequence, the
interest in the subject is reduced and the production of good physics
teachers in the future will become more difficult. Most of the small
colleges have squeezed their physics programs and reduced faculty size. But,
the others have made physics a requirement for more programs, to keep a
better level of physics enrollment. However, this can help the undergraduate
programs only. Attracting more students to physics majors is an even bigger
challenge.

At this point, we cannot ignore the fact that a significant percentage of
teachers are not able to play a role in the improvement of standards and /or
attracting more students to physics and change this image of the subject
itself. It is not because they are not competent enough to do that or they
do not care about the promotion of the subject. It is mainly because they do
not have job security. A large proportion of teachers have a great
limitation in playing a big role in attracting students to the subject.
There is no room to try new techniques or policies to increase standards and
take the risk of getting students less happy with teachers. It will always
lead to the insecurity of a teacher's job. After all, they need money for
their survival. New teachers would have to care more about the student's
evaluation than the standards, anyway. When teachers have to compromise on
standards for their job security, their dedication level is obviously
affected by it. When their job commitment is not acknowledged at all and the
financial pressures and insecurity leads them to the frustration level where
they are badly discouraged to work for standards and / or to contribute good
research. We need to discuss in detail the role of teachers to justify that
the teacher's frustration is not only for them, it has the long-lasting
effects on future development. Anyway, we discuss some of the major factors
behind this image of physics, the limitations produced by it and the
suggestions to improve the image of physics among students. It is an even
much more complicated issue and needs to be discussed thoroughly. We can
briefly summarize these problems as follows:

\begin{enumerate}
\item  Lack of appropriate math background among students.

\item  Decreasing ratio of serious students for the sake of knowledge and
growth in the easy-going approach among students to get a degree for better
jobs and salaries, quickly.

\item  Shortage of outside the classroom time for students.

\item  Lack of understanding of the subject from the beginning.

\item  Pressure on the contract teachers because of the security of teaching
jobs that greatly depends on the student's evaluation.
\end{enumerate}

All of these points could separately be a research topic in itself. We have
also postponed the discussion of the effect of this adjunct culture (due to
the presence of part-time teaching faculty) on physics research. There are
many adjuncts that are as specially interested in research and could
contribute to good physics research if some job security and peace of mind
is provided to them as the necessary components for the original research.
We have seen adjuncts who would still like to do some research after
teaching for something like twenty hours in a week. However, their research
efforts are not even acknowledged by the institutes where they are teaching
as adjuncts. Though they will use the college addresses and all that. In
addition to job insecurity, the rules for research grants also need to be
revised for the promotion of research in the country. We will come back to
the details of this issue somewhere else. However, we will give here some of
the analysis based on our long-time involvement in physics teaching and our
tremendous interest in the promotion of physics in developing countries. In
these countries, limited resources and financial issues are bigger obstacles
in higher studies, in general and physics education, in particular.
Apparently these problems are not limited to one college course only. They
have much deeper roots and need more detailed investigation to reach the
root cause of the problem. We will look at some of the immediate issues
individually in the following sections.

\section{MATHEMATICAL BACKGROUND}

Mathematics is a very important tool to learn and understand physics.
Difficult physics problems can be resolved, with equations and their
physical interpretation. A good math background is always needed for physics
because the mathematical equations always help in understanding the physical
concepts. So, math courses are prerequisites for the physics courses. These
prerequisites are an additional problem for physics enrollment, but we
cannot do much about it because some math is needed for understanding
physics. On the other hand, most of the students cannot develop appropriate
math skill because they do not have enough background. Also their aptitude
for math learning and the level of math teaching varies from college to
college and even from teacher to teacher. There are other similar types of
problems in the math courses, which we will briefly discuss here.

It is almost natural to think, in this computer age, that we do not need to
spend a lot of time on learning and practicing math if we can easily use
computers to get mistake-free reliable results. Computers can easily do
almost all the mathematical calculations for us. The only thing needed is
computer proficiency, which almost all modern students have, and are getting
better in time. There is no problem in stating that their use makes research
life easier and mistake-free and also saves time. Nobody can disagree with
the fact that computation becomes a lot more reliable and quicker and even
more detailed with computers. However, elementary physics needs a little
more understanding of math to start with. Some hand calculation skill is
needed to go through the derivation of physics equations and also the
interpretation. Students need some practice in simple math to at least get
used to playing with numbers. But calculators are usually introduced at the
elementary level; this does not permit the students to understand very basic
ideas of arithmetic. This easygoing approach in math helps to generate math
fear because they never get used to the calculations, properly. It has also
been noticed that there is not so much a lack of math aptitude, as it is the
lack of time and dedication to use math. This is because students are
introduced to the easier way, using calculators and computers at very early
stages of their student life. We would recommend encouraging school kids to
do more hand calculations and to play more with numbers by hand in the
elementary schools. Kids could be taught computers for other purposes but
should be discouraged to use calculators for many calculations, even in the
middle schools. Math skill is definitely affected by the basic training.
This way, students will have enough skill to play with numbers and will be
more used to hand calculations. Thereby, math teaching could become easier
at higher levels.

Another big reason for lack of math interest and even math fear is linked
with the problems of teachers. The performance of the teachers is usually
judged in the light of students grades at all levels. So, to show better
grades of the students for their own job security, they have to adopt
examining and grading systems that can help students get good grades. In
this process, sometimes, there can be compromises on the standards and
loopholes are created in their knowledge and practice. This is obviously
dependent on the individual school and the teacher. Again, high school
standards are not our topic at the moment but its impact on college
educations is inevitable. These problems have some long-lasting effects on
the high school graduates who are college freshmen. As a result, students
are not usually getting appropriate math training before entering the
colleges. High school teachers are under the pressure of getting a high rate
of graduation and have to keep it high for their job security. So they have
to compromise on standards sometimes. Students usually have very limited
practice. Study of sorted topics and exam preparation is the usual attitude
of students. They use short-cut methods to get good grades even without
adequate knowledge and the required practice for higher-level education. In
short:

\begin{enumerate}
\item  Lack of practice in math does not let the students develop enough
interest in advanced math, even if they have enough interest to start with

\item  Financial responsibilities including financial arrangements for their
college education. Shortage of time outside the classroom and several other
factors are additional issues.

\item  Students have to study a number of courses at the same time. So they
do not have enough time for each subject. In this situation finding extra
time for the math practice is not possible.
\end{enumerate}

In this situation, math background plays a big role in the declining trend
of physics enrollment. The detailed discussion of declining math background
is a complete research topic in itself. Physicists in the field are very
well aware of this problem and try to minimize its impact on the standards
of physics courses. There are two usual approaches to handle this issue.
Some of the teachers try to assume minimum math background and go over the
mathematical steps in detail to make things clearly understandable to
physics students. Others try to reduce math in their elementary level
courses and try to explain the physical concepts using examples from daily
life and certain analogies. Sometimes both of these methods are used to
overcome math deficiency. It depends on the personal judgment of teachers
about their class on how weak math background can be treated properly.
Sometimes if this weakness is limited to a few students, then individual
attention to those students and/or tutoring through learning centers in
colleges can give enough help. But this could only be helpful if students
are willing to get help from them and have a strong urge to study physics.
This seems to be lacking these days. However, along with all the individual
and collective efforts made by teachers in this direction some of the highly
contributing problems are listed in the next section.

\section{COMMON PROBLEMS OF PHYSICS TEACHERS}

Now we come to a comparatively sensitive issue in the sense that some of us
may not agree with it at all or may strongly disagree at certain points.
However, we are trying to address this issue honestly and sincerely because
when it comes to the analysis, it should be genuine and very honest. Our
point of view may be different from others but our analysis is fair as it is
based on our experience and on study of the issue in detail. And I think
that we as physicists want to promote physics and want to see it as a
popular subject. All of our discussions here are for the sake of this
purpose.

Physics is a subject that can only be taught effectively with full
dedication. This dedication is only possible with job security and financial
satisfaction. All teachers do not have this now. With financial worries and
job issues, it becomes very difficult to concentrate on standards rather
than concerning oneself about evaluations by students in order to get a new
teaching contract for the new semester or the new quarter. Tenured teachers
are proved to be much better teachers due to job security and financial
stability. They also have much more liberty to choose courses of their own
interest and introduce new courses. They can pay more attention to teaching
standards and try new effective tools on students for better results. They
obviously cannot afford to do extensive surveys or work on the long research
projects including the research in physics education.

On the other hand, contract teachers cannot afford this. However, even among
the tenured teachers, everyone does not have equal interest in physics
education, obviously. But almost 20-30\% teachers and sometimes even more
are on quarterly/monthly or yearly contracts. The continuation of their job
is not usually based on the quality of their teaching or how much they care
about the effective teaching. These teachers have extensions in their
contracts depending on the evaluation by students. This instructor
evaluation is apparently the only way to check the teaching ability of
contract teachers. If, due to some reason they take more interest in
standards and do not care about the evaluations, they may be sent home, even
if they demonstrate their dedication to students and their role in effective
teaching is unquestionable, otherwise. They do not even have options to pick
up courses of their own choice and show their real teaching abilities and
their professional strength. The opportunity is missed to show their full
abilities or their own teaching style, due to the risk of losing their
teaching assignments. They usually have to teach in the presence of mentors
or tenured teachers and are expected to follow the styles and standards of
their mentors. They have no way to introduce/try new ways to improve the
quality. Their abilities are not tested even if they have more experience
and knowledge. It is just their status that is counted. In a way, their
contributions to the improvement of standards is not even tested or judged.
All they need is good instructor evaluations, which depends on several other
factors also.

Usually, the majority of students do not like teachers who are more
demanding and are committed to raise standards. The student's background and
their seriousness in study and the expectations from the course itself are
some of the additional factors. There are several unknown parameters for new
teachers, which can change evaluations altogether, regardless of the
teaching quality or effectiveness. So the method of selection of contract
teachers does not help much in effective teaching. Schools save money in
hiring contract teachers instead of full-time visiting positions, at the
cost of the standards of education. They can pay half of the minimum salary
without any employment benefits to contract teachers and ignore the
standards. Contract teachers can be acceptable for several subjects, but
physics has a different situation. So these less expensive contract
positions are affecting the standards of physics teaching badly, not because
colleges do not hire equally qualified teachers, it is because they cannot
get best out of their abilities.

Another issue associated with contract positions is the availability of
teachers to students outside the classroom. This especially occurs when
adjunct instructors come to the campus for a very limited time. They do not
have proper offices and cannot usually be available other than at the class
time. If they give a couple of office hours for their own satisfaction, most
of the students cannot see them in this given limited time because of the
their schedule and most of the colleges do not require office hours for
adjuncts because they do not want to pay them for this. Most of the adjunct
faculty has to have some other source of income in addition to teaching so
their involvement is not at the same level as that of the full-time faculty.
Some colleges offer students help from other teachers, which may not be
equally helpful either. Moreover, since adjunct positions are not long-term
positions so adjuncts cannot have longer term planning with students for
next term courses or undergrad research involvement and so on. Because of
their minimum exposure to their teaching departments, they do not attend any
meetings or understand very well the requirements of programs for which they
are teaching. Again the simple solution to all these issues is to replace
adjunct positions with full time short term contract positions at least, so
that the part time faculty can work in one college at a time and spend some
more time with students and teach more effectively and peacefully. They
could be given more workload to compensate the salaries, so that they do not
have to drive from one place to another place, which definitely affect the
quality of teaching. Adjunct teachers can easily make an amount of money
that is comparable to the regular starting salary of an undergraduate level
teacher with some additional workload. The only problem is the benefit
package that could be available for the adjunct positions in special
situations also. So instead of distributing this money among different
colleges, some better arrangements could be done.

Some of the other common job problems for the qualified physics teachers are
briefly unveiled here. The physics job market is normally low. There are
several reasons behind this. Low enrollment of physics students is a big
factor in reduced teaching positions. Teaching jobs are not well-paid jobs
anyway. On the other hand, physicists are not much demanded in other jobs
either. Especially, there are not so many openings in the field of
theoretical physics. Usually, physicists have to depend on their side skills
for livelihood and not their main expertise in physics. So the bright
students would always be attracted to those programs which are most demanded
for better paying jobs. On the other hand, we cannot deny the fact that
physics majors have to be hard-working students with good mathematics
background. Especially for theoretical physics, a strong math background and
full interest in the subject with full commitment is needed. So it was a
growing field when brilliant students chose physics because of the better
job opportunities. The low enrollment in physics courses and in physics
program lead to job cuts in physics everywhere and physics programs
squeezed. The other factor leading to the low enrollment is the funding
mechanism. The deficiency in basic theoretical physics research funds is a
big discouragement for attracting brilliant students to adopt physics as a
career. So physics is not an attractive career any more.

\section{OTHER ISSUES}

Some of the other important issues contributing to our topics are related
directly or indirectly to social and economic issues and affect education in
general. One of the big problems, which students have to face during
studies, is related to the economic matters. Usually, college students have
to take care of all their financial needs. For that purpose they have to
work during studies. Part-time jobs are still manageable, but some of the
students have to work a lot to fulfill their financial needs, which includes
college spending also. Of course, they take this out of their study time and
hence prefer to take the subjects that do not need a lot of
out-of-the-classroom time. Sometimes, they have to take several courses to
make their study less expensive and save on college fees for a semester. In
that case, they have a stipulated period of time for every subject. In this
situation, they are not usually able to fit physics in their schedule and if
they do, they cannot do well.

Furthermore, most of the elementary level physics classes are comprised of a
very diverse academic background and include different majors. These
students come from all different places. It means that even the same
prerequisites indicate different academic levels, depending on their
previous institute and teachers. And, it is even more complicated in
colleges. They come from different majors including all science and liberal
arts majors, and even have spent different numbers of years in college. They
have different reasons to enroll in physics courses. Some of them like
physics and may even end up selecting physics as a career, whereas others
would be sitting there just to complete the required credits and not to
learn physics. In this case, it is really very difficult to teach large
classes. This means if you are teaching a couple of hundreds of students you
can always make four or five classes out of them with quite different
backgrounds and different course goals.

In this situation a teacher cannot equally satisfy all the students at the
same time. The level of the course has to be kept at the point where most of
the students can pass it. This is how we run into the risk of losing some of
the interested physics students who could have continued if they were taught
in a different way. They could have put in more individual time to develop a
better understanding of the subject. These types of students need a little
bit more one-on-one discussion time with teachers and need to have more
questions asked in class. It is impossible to satisfy them in big classes.
Smaller class size could be very helpful in physics teaching because a
teacher can pay individual attention to the students. Knowing that there are
different reasons for the lack of interest in different types of physics at
different levels, we should try to resolve these issues, in general.

Physics is being taught in high schools and colleges at different levels. We
can classify these courses as the elementary, medium and higher-level
courses. The students have their first exposure to compulsory physics
courses in high schools. Compulsory physics courses at the college level
come afterwards. These courses are either based on algebra or simple
calculus. They are offered to big groups of students with a great variety of
background and interest. In this situation, it becomes very difficult for
teachers to make the subject attractive enough for the students to prefer
physics to other better paying, easy-to-understand and seemingly more
relevant subjects.

Once the students choose physics as a major subject, then their interest is
clear. Now, just to keep the students interested in the subject, the physics
courses have to be kept parallel (in time consumption) to the courses in
other disciplines. So, class demonstrations and problem solving strategies
are used to help them learn the topics quickly and properly understand the
concepts. This is very effective in developing working knowledge of the
basic physics concepts among students of the engineering departments.
However, in our point of view, deeper understanding of physics requires a
combination of the observational/experimental as well as the mathematical
approach, even at the elementary level. Sufficient theoretical background is
needed to develop a real interest in the subject. For the rest of their
career they can easily learn and understand through the physical
demonstrations and the problem solving techniques.

Physics teaching is also done in engineering departments to give engineers
the relevant physics background. Usually, the emphasis is on working
knowledge and providing sufficient practice to use the relevant equations to
solve given problems. This is understandable in that the main objective is
to create acceptable and safe designs and resolve engineering problems.
However, I would think that good engineers should also have a better
background in the relevant areas of physics. This will help them throughout
their career to face any unforeseen situation and make innovative designs.
For this purpose, the engineering departments can arrange a couple of
courses to develop this background. Good physics background with hands-on
experience is equally helpful in medicine, experimental chemistry and
biology, especially, when it comes to the use of optical, electronic or
mechanical equipment. So all the science majors should have reasonable
exposure to physics to develop better understanding in their own subjects
also. Here, I would like to mention a personal experience. I asked about 60
students of Mechanical Engineering to define the word `mechanics'. Nobody
could answer this question. Then I asked, what is Mechanical Engineering?
And the answer was!! Problem Solving. We would expect them to at least know
the role of mechanics in building modern inventive mechanical designs.

The next issue is funding. Since undergraduate research funds are mainly
allocated for the experimental or applied science projects, new students are
encouraged to choose their major research projects in experimental/applied
physics areas. It is an established fact that there are several theoretical
physicists from developing countries, which could convert from theoretical
to experimental physics research after working for some time with the
experimental research groups. However, a lot of work is required to develop
enough mathematical background to participate in theoretical research. In
recent times, any good experimental group requires computational skills and
theoretical physicists with good mathematical and physical background could
serve this need and become an irreplaceable part of an experimental group.
They could even replace computer engineers sometimes, because of their
insight in physics.

High-level physics teaching is not our topic at the moment, because when the
students reach that level, they already have sufficient interest and
learning capabilities to cope with the complexities of the subject.
Therefore, the student's response and interest in the subject makes teaching
enjoyable and academically challenging for the teachers. But that interest
is only developed under favorable circumstances. At that stage, subject
knowledge of the teachers is appreciated more than their handling of
students or teaching styles.

\section{CONCLUDING REMARKS}

In this section we will like to conclude our discussion by briefly
mentioning what we think about this issue. It is clear that the decrease in
physics enrollment should be taken seriously and we should try to take care
of this problem in the light of all the above available details. It might
lead to very serious consequences if it was not controlled at this stage.
Some of our concerns are clear from the available statistical data [6].
However, we would like to emphasize a few points that include ways to
re-establish the physics departments and to initiate undergraduate majors in
two-year colleges. In this time of availability of more qualified physicists
than suitable jobs, two-year colleges can also hire more qualified
physicists who can participate in raising educational standards in such
institutes. If the teaching load could be lowered to some extent and
research is at least not discouraged, it will also lead to improvement in
the educational standards. This strategy will attract serious researchers
who want to participate in the improvement of physics education, also.
Instructor evaluations are very important but they should be more used to
help committed teachers to determine the effectiveness of their teaching
methods and should not be used as the sole criteria for the teacher's job
extension.

We will even recommend at least full-time temporary hiring instead of the
adjunct position. Lab teaching could be handled by adjuncts but the
full-time teachers should preferably teach lectures so that they can give
enough time to students. In that case lab teaching can also be done by
full-time faculty and could be more organized and effective to attract more
students and improve the image of the subject itself. Moreover, better ways
for the instructor evaluation procedure could be developed incorporating the
student's interest in the subject in the evaluation procedures. We are
already aware of the fact that some colleges have very good evaluation
procedure and the obvious difference in the teaching standards could be
easily observed. Another convenience could be achieved by introducing a
comparatively uniform procedure of instructor evaluation so that the
instructor evaluations in different institutes could be compared. Some
further research is required in this direction. Undergraduate physics
research programs are the effective ways to promote physics among students.

\section{RECOMMENDATIONS}

In the end, we would like to give a few suggestions to secure our future
standards in technology and science. These are obviously some of the
possible moves in the direction of the promotion of physics among the
incoming generation. We know that these are not very easy steps, especially
to take all at once. We understand the limitations of different policy
making bodies and are aware of the financial constraints. However, some of
the improvements are possible partially or completely. Consideration of
these issues will at least help to protect us from the further decline in
standards of physics education. This is the first needed move. We know some
of these things are already adopted by certain places, however we have to
mention them here to suggest it to everybody. Some of the recommendations
are as follows:

\begin{enumerate}
\item  Promotion of science interest in elementary schools, in particular by
developing calculation skill in math. We would recommend completely banning
the use of calculators in elementary schools. Even at higher-level school
students should be discouraged to use calculators for simple numerical
calculations.

\item  School trips can be organized to some research labs and students
could be provided with opportunities to meet renowned scientists of the
area. Some other programs can be designed to involve college/university
teachers to work with school students.

\item  Lesson the pressure for a high graduation rate. So, job security and
promotion of teachers should be less dependent on a high graduation rate.
They should be required to involve students in more practice in solving
problems. State exams could be designed to accommodate the testing of
student's math practice and their broad knowledge.

\item  In all high schools, those teachers who have degrees in the relevant
subject should be teaching math and science courses. Preferably, M.S degree
in the relevant subject and should be a requirement. Some of the schools
already have these standards.

\item  Community colleges and state colleges should give more full-time
positions instead of hiring adjuncts for teaching those courses. So many
highly qualified people are looking for physics jobs these days that it
should not be difficult to hire one teacher who can teach for example every
undergraduate course. Physics degree holders should preferably, do physics
teaching. Preference should be given to doctorate degree holders.

\item  Instructor evaluation by students should be considered as input data
for further improvement in teaching and not the sole criteria for job
extension.

\item  The teaching load in colleges should be reduced to improve the
quality of teaching. Teachers should have enough to organize lectures and
design experiments properly.

\item  Qualified technicians should also be provided to run physics labs
properly. These technicians can take care of the equipment, which are never
used by many colleges, just because teachers have no time. This expensive
equipment could be used to introduce new experiments.

\item  Research should not be discouraged in colleges at all. Actually
college teachers should be encouraged to establish research collaboration
with universities. In this way, they will also be more aware of the learning
needs of students who want to transfer to other colleges to do research in
basic sciences. They can play their role to promote basic sciences by
getting new ideas and having discussions with their colleagues in research
institutes.

\item  Physics courses for engineers should be designed more clearly, so as
to relate physics to engineering applications. Students should be able to
learn and understand physical concepts so that they can properly apply them
in their engineering courses.

\item  Research grants should be given to contract teachers by colleges to
involve undergraduate students in research.

\item  More inter-disciplinary courses should be designed to make students
more familiar with physics and other science courses. Especially, the
courses, which can link natural sciences and social science together, could
be very interesting. For this purpose, teachers with more than one master
degree could be very helpful.
\end{enumerate}

ACKNOWLEDGMENTS: The author would like to thank Prof. Alan Miller of the
Physics Department of Syracuse University to read the manuscript and to give
very helpful suggestions.

\begin{center}
REFERENCES
\end{center}

\begin{enumerate}
\item  Milton W. Cole and Carla Zembal Saul, `Enhancing Science Education in
Elementary Schools`, arXiv: physics/0207051 (to appear in American Journal
of Physics)(2002)

\item  Karl Svozil, `Peer Review in Context`, arXiv: physics/0208046 (2002)

\item  N.D.Finkelstein, `Teaching and Learning Physics` arXiv:
physics/0505091(2005)

\item  Emily M.Reiser and Mark E. Markes, `An inferential analysis of the
effect of activity based physics instruction on the persistent
misconceptions of lecture students` arXiv: physics/0506208(2005)

\item  William J.Gerace and Ian D. Beatty, `Teaching vs.learning: Changing
perspectives on problem solving in physics instructions` arXiv:
physics/0508131(2005)

\item  Roman Czujko, www.aip.org/statistics/trends/reports/june9talk.pdf 

\item  Richard R. Hake, Am. J. of Physics 66, 64 (1998)

\item  Extensive surveys like the one in Ref. 7 and several other articles
on similar type of subjects
\end{enumerate}

\end{document}